\documentclass{article}

\usepackage{PRIMEarxiv}
 
\usepackage{amsmath,amssymb,amsfonts}
\usepackage{array}
\usepackage{url}
\usepackage{orcidlink}
\usepackage{tabularx}
\usepackage{multirow}
\usepackage{booktabs}    
\usepackage{makecell}  

\usepackage{algorithm}  
\usepackage{algorithmic} 

\usepackage[utf8]{inputenc} 
\usepackage[T1]{fontenc}    
\usepackage{hyperref}      
\usepackage{url}          
\usepackage{booktabs}      
\usepackage{amsfonts}     
\usepackage{nicefrac}     
\usepackage{microtype}  
\usepackage{lipsum}
\usepackage{fancyhdr}    
\usepackage{graphicx}      
\graphicspath{{media/}}    
\usepackage{authblk}

\title{TGSD: Topology-Guided State-Space Diffusion Framework for EEG Spatial Super-Resolution}

\author[1,2]{Zijian Kang}
\author[2,$\star$]{Weiming Zeng}
\author[1,2]{Yueyang Li}
\author[2]{Shengyu Gong}
\author[3]{Hongjie Yan}
\author[1]{Wai Ting Siok}
\author[1,$\star$]{Nizhuan Wang}

\affil[1]{\textit{Lab of Digital Image and Intelligent Computation, Shanghai Maritime University}, Shanghai 201306, China}
\affil[2]{\textit{Department of Department of Language Science and Technology, The Hong Kong Polytechnic University}, Hong Kong SAR, China}
\affil[3]{\textit{Affiliated Lianyungang Hospital of Xuzhou Medical University}, Lianyungang 222002, China}
\affil[$\star$]{Correspondence: wangnizhuan1120@gmail.com; zengwm86@163.com}

\begin{document}
\maketitle

\begin{abstract}
Low-density EEG is more suitable for wearable and IoT-based brain sensing, but sparse electrode sampling often lacks sufficient spatial information to characterize cross-regional neural activity. EEG spatial super-resolution aims to recover dense-channel EEG from sparse recordings, yet remains challenging because channel missingness typically occurs at the whole-channel level, spatiotemporal dependencies over the full electrode layout are often underexplored, and the mapping from sparse to dense signals is inherently ambiguous. To address these issues, we propose TGSD, a topology-guided state-space diffusion framework for EEG spatial super-resolution. TGSD first employs a Hierarchical Spatial Prior Encoder to learn topology-aware priors over the complete electrode layout by integrating local geometric relationships with region-level contextual information. Based on these priors and sparse observations, a Conditional State-Space Diffusion Reconstructor progressively generates missing-channel signals through reverse diffusion, while alternating temporal and channel-wise state-space modeling captures long-range temporal dynamics and inter-channel dependencies in a unified framework. Experiments on the SEED and PhysioNet MM/I datasets show that TGSD consistently outperforms representative baselines under different super-resolution factors in both reconstruction fidelity and downstream classification performance. These results demonstrate the effectiveness of combining topology-aware spatial priors with conditional diffusion for enhancing practical low-density EEG sensing in wearable and IoT scenarios. The official implementation code is available at \url{https://github.com/jtggz/TGSD}.
\end{abstract}


\section{Introduction}
Electroencephalography (EEG) has long been used in clinical diagnosis and neuroscience research\cite{biasiucci2019electroencephalography}, and is increasingly being adopted in wearable and Internet of Things (IoT) settings because of its noninvasiveness, low cost, and high temporal resolution. This shift enables a wide range of daily-life brain sensing applications, such as continuous emotion and stress monitoring to support mental health management, concentration enhancement in sports and meditation, and speech/language decoding\cite{zhang2026linguistics,s26103212, jenke2014feature,xu2018review}. Although conventional high-density EEG systems provide rich spatial sampling, they are often unsuitable for these applications due to their bulky equipment, complex preparation, limited wearing comfort, and higher costs\cite{lau2012many}, whereas wearable low-density devices are more portable and deployment-friendly yet suffer from sparse spatial coverage\cite{hinrichs2020comparison,li2025tale}. This trade-off makes it difficult for practical low-density EEG systems to reliably capture cross-regional neural dynamics, motivating the study of EEG spatial super-resolution.

EEG spatial super-resolution aims to reconstruct dense-channel EEG from sparse channel observations, thereby bridging the gap between practical low-density acquisition and high-density spatial representation\cite{shin2024super}. Unlike image super-resolution on regular grids\cite{wang2020deep}, this task requires recovering multichannel temporal signals over an irregular electrode layout. Moreover, missingness typically occurs at the whole-channel level rather than at isolated time points, making the problem a structured spatiotemporal reconstruction task instead of a conventional interpolation problem. The goal is therefore not only to infer plausible signals at unobserved electrode sites, but also to preserve temporal dynamics and spatial consistency across the entire electrode layout. 

Existing studies have explored several directions for EEG channel reconstruction, including spatial interpolation, discriminative deep regression, and topology-aware neural modeling. Interpolation-based methods are simple and efficient, but they are usually built on fixed geometric assumptions and local smoothness priors, which limits their ability to capture complex nonlinear spatial variations\cite{courellis2016eeg}. Deep regression models improve representation capacity, yet they typically learn a deterministic mapping from sparse observations to dense signals, which tends to produce over-smoothed reconstructions when the underlying signal distribution is ambiguous\cite{kwon2019super}. Graph- or topology-based methods introduce structural information among electrodes, but many of them are designed for classification or decoding tasks, or construct dependencies only over observed channels, making them less suitable for reconstructing missing channels under the full electrode layout\cite{song2018eeg}. More importantly, relatively few methods jointly address three issues central to EEG spatial super-resolution: full-layout topology modeling, uncertainty-aware reconstruction, and unified temporal--channel dependency learning.

To address these limitations, we propose TGSD, a Topology-Guided State-Space Diffusion framework for EEG spatial super-resolution. TGSD consists of two key components. First, a Hierarchical Spatial Prior Encoder (HSPE) learns topology-aware priors over the complete electrode layout by integrating local geometric interactions with region-level contextual aggregation, thereby embedding both observed and target channels within a shared spatial representation. Second, a Conditional State-Space Diffusion Reconstructor (CSDR) generates missing-channel EEG through conditional reverse diffusion guided by sparse observations and the learned spatial prior. Within the denoising network, alternating temporal- and channel-wise state-space modeling is adopted to capture long-range temporal dependencies and cross-channel interactions efficiently. In this way, TGSD integrates spatial topology modeling, probabilistic reconstruction, and temporal-channel dependency learning into a unified framework for recovering high-density EEG from low-density measurements, highlighting its potential for low-density wearable EEG and IoT-based neural sensing applications.

\begin{figure*}
    \centering
    \includegraphics[width = 1\textwidth]{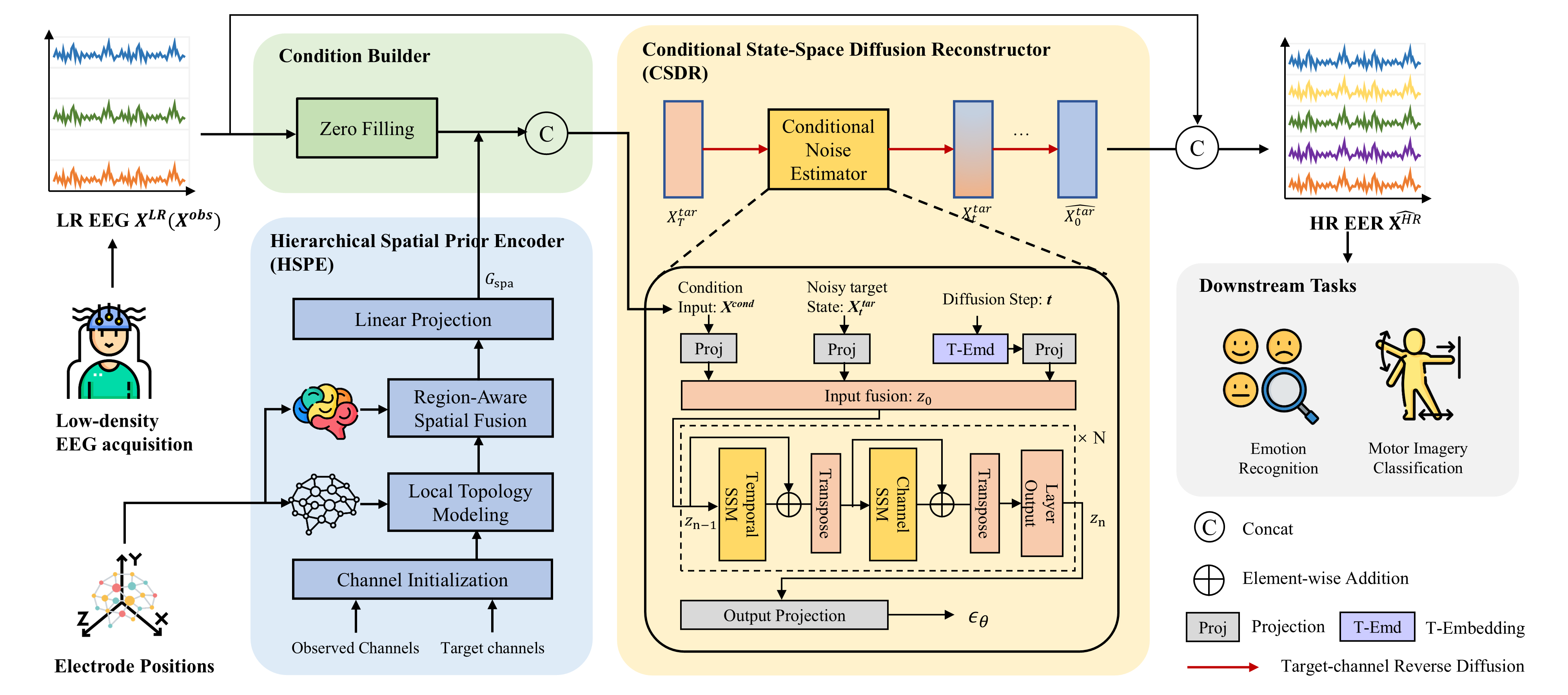}
    \caption{Overall framework of TGSD for EEG spatial super-resolution. This framework consists of a Hierarchical Spatial Prior Encoder (HSPE) for topology-aware spatial prior learning and a Conditional State-Space Diffusion Reconstructor (CSDR) for target-channel EEG reconstruction via reverse diffusion.}
    \label{fig:framework}
\end{figure*}

\section{Related Work}
\label{title:Related Work}

\subsection{EEG Spatial Super-Resolution and Missing-Channel Reconstruction}

EEG spatial reconstruction has been studied to alleviate the loss of spatial information caused by low-density acquisition. Early approaches mainly relied on interpolation-based techniques, such as spherical spline interpolation and its dynamic variants, to estimate signals at unobserved electrodes \cite{freeden1984spherical,khouaja2016enhancing}. These methods are simple and do not require large-scale training data, but they usually depend on fixed geometric assumptions and local smoothness priors, which limits their ability to capture complex nonlinear spatial patterns in EEG activity.

More recent studies have introduced data-driven models for EEG channel reconstruction. Convolutional networks have shown the feasibility of learning sparse-to-dense mappings from data \cite{han2018feasibility,kwon2019super}, while generative adversarial learning has also been explored for missing-channel recovery \cite{corley2018deep}. Transformer-based models further improved spatiotemporal modeling capacity for EEG super-resolution \cite{li2025estformer}, and graph-guided methods incorporated structural or functional connectivity priors to enhance reconstruction quality \cite{tang2022deep}. Other efforts have considered virtual electrode reconstruction and matrix-decomposition-based recovery \cite{svantesson2021virtual,tang2025eeg}. Despite this progress, most existing methods either rely on deterministic regression or make limited use of structural priors over the complete electrode layout, which restricts their ability to model reconstruction uncertainty and provide sufficient context for missing channels.

\subsection{Topology-Aware EEG Modeling}

Topology-aware modeling has become an important direction in EEG analysis because inter-channel dependencies are closely related to scalp geometry, brain-region organization, and functional connectivity. Existing studies have widely represented EEG channels as graph nodes and employed graph convolution, graph attention, or adaptive graph learning to capture spatial interactions among electrodes \cite{cai2023brain,li2023effective}. Beyond single-scale graph construction, hierarchical and multi-scale schemes have also been explored to model local and long-range dependencies over the scalp \cite{jin2024pgcn}. These studies demonstrate that explicitly incorporating topology can substantially improve EEG representation learning.

However, most topology-aware EEG models are developed for classification, recognition, or decoding tasks rather than channel reconstruction. In addition, many methods construct spatial dependencies only among observed channels or use a single-scale topology, making them less suitable for super-resolution settings where target channels are entirely missing. Therefore, EEG spatial super-resolution still requires a spatial encoding mechanism that can learn multi-level priors over the full electrode layout and provide topology-aware context for both observed and missing channels before reconstruction.

\subsection{Diffusion-Based Generative Reconstruction and Sequence Modeling}

Diffusion models have recently shown strong performance in probabilistic generation and uncertainty-aware reconstruction. Compared with deterministic regression, diffusion-based methods are better suited to one-to-many inverse problems because they model the target distribution through iterative denoising. In time-series modeling, diffusion models have been applied to sequence generation, imputation, and spatiotemporal recovery \cite{yuan2024diffusion,liu2025rdpi}, and have also been extended to neural signal generation and restoration \cite{vetter2024generating}. For EEG spatial super-resolution, recent studies have begun to adopt diffusion-based reconstruction frameworks \cite{wang2025generative,liu2025step}, which highlights the potential of generative modeling for recovering missing EEG channels under sparse observations.

At the same time, EEG reconstruction requires an efficient backbone for modeling long temporal sequences and cross-channel dependencies. Compared with recurrent models and Transformers, state-space models (SSMs) offer a favorable balance between long-range dependency modeling and computational efficiency \cite{hamilton1994state,gu2021efficiently,dao2024transformers}. Recent studies have further combined SSMs with diffusion frameworks for time-series generation and imputation \cite{gao2025ssd}, suggesting their promise as denoising backbones for structured signal reconstruction. Nevertheless, existing EEG spatial super-resolution methods still make limited use of explicit topology-aware priors over the full electrode layout, and unified modeling of probabilistic reconstruction with both temporal and channel dependencies remains underexplored. To address these limitations, we propose TGSD, which integrates topology-guided spatial prior learning, conditional diffusion-based reconstruction, and temporal-channel state-space modeling within a unified framework.

\section{Method}
\label{title:method}

As illustrated in Fig.~\ref{fig:framework}, TGSD comprises a Hierarchical Spatial Prior Encoder (HSPE) and a Conditional State-Space Diffusion Reconstructor (CSDR). HSPE learns a topology-aware prior over the full electrode layout from sparse observations and electrode coordinates, while CSDR reconstructs the missing-channel EEG through conditional reverse diffusion guided by the learned prior.

\subsection{Problem Formulation}

Let an EEG segment be denoted by $\mathbf{X}\in\mathbb{R}^{K\times L}$, where $K$ is the total number of electrodes and $L$ is the sequence length. For a spatial upsampling factor $s\in\{2,4,8\}$, the full electrode set is partitioned into an observed channel set $\mathcal{C}_{obs}$ and a target channel set $\mathcal{C}_{tar}$, with $K_{obs}=\lfloor K/s \rfloor$ and $K_{tar}=K-K_{obs}$. The corresponding observed and target EEG are denoted by $\mathbf{X}^{obs}\in\mathbb{R}^{K_{obs}\times L}$ and $\mathbf{X}^{tar}\in\mathbb{R}^{K_{tar}\times L}$, respectively.

Let $\mathbf{P}\in\mathbb{R}^{K\times 3}$ denote the 3D coordinates of all electrodes. The goal of EEG spatial super-resolution is to reconstruct the target-channel signals from sparse observations and electrode positions:
\begin{equation}
\hat{\mathbf{X}}^{tar}=\mathcal{F}_{\theta}(\mathbf{X}^{obs},\mathbf{P}).
\end{equation}
Since multiple plausible high-density patterns may correspond to the same sparse observation, we model this task as conditional generation by learning $p(\mathbf{X}^{tar}\mid \mathbf{X}^{obs},\mathbf{P})$.

To reflect practical low-density EEG acquisition, we consider channel-level missingness rather than temporal point-wise masking. The mask matrix $\mathbf{M}\in\{0,1\}^{K\times L}$ is defined as
\begin{equation}
M_{i,j}=
\begin{cases}
1, & i\in\mathcal{C}_{tar},\\
0, & i\in\mathcal{C}_{obs}.
\end{cases}
\end{equation}
That is, once a channel belongs to $\mathcal{C}_{tar}$, all its temporal samples are missing in the current segment.

\subsection{Topology-Guided Spatial Prior Encoding}

HSPE learns a spatial prior $\mathbf{G}_{spa}\in\mathbb{R}^{K\times F_{spa}}$ over the complete electrode layout. Unlike methods that encode only observed channels, HSPE explicitly includes both observed and target channels, allowing missing electrodes to acquire structural context before reconstruction.

We first initialize a latent representation for each channel. For observed channels, the input EEG sequence is projected into the latent space, while target channels are initialized by learnable embeddings:
\begin{equation}
\mathbf{h}_{i}^{(0)}=
\begin{cases}
\phi(\mathbf{x}_{i}), & i\in\mathcal{C}_{obs},\\
\mathbf{e}_{i}, & i\in\mathcal{C}_{tar},
\end{cases}
\end{equation}
where $\phi(\cdot)$ is a learnable sequence projection and $\mathbf{e}_{i}\in\mathbb{R}^{F_h}$ is the embedding of the $i$-th target channel.

We then model local geometric relations based on electrode coordinates. Let
\begin{equation}
d_{ij}=\|\mathbf{p}_{i}-\mathbf{p}_{j}\|_{2}
\end{equation}
be the Euclidean distance between electrodes $i$ and $j$. The local adjacency is defined as
\begin{equation}
A^{loc}_{ij}
=
\exp\!\left(-\frac{d_{ij}^{2}}{\tau^{2}}\right)\cdot \mathbb{I}(j\in\mathcal{N}_{k}(i)),
\end{equation}
where $\tau$ controls spatial decay and $\mathcal{N}_{k}(i)$ denotes the $k$ nearest neighbors of node $i$. Based on the normalized adjacency $\tilde{\mathbf{A}}^{loc}$, local topology propagation is performed as
\begin{equation}
\mathbf{H}^{loc}
=
\sigma\!\left(\tilde{\mathbf{A}}^{loc}\mathbf{H}^{(0)}\mathbf{W}_{loc}\right),
\end{equation}
where $\mathbf{W}_{loc}\in\mathbb{R}^{F_h\times F_h}$ is learnable and $\sigma(\cdot)$ is a nonlinear activation.

To capture broader scalp organization, HSPE further introduces region-aware contextual fusion. Specifically, the electrode layout is partitioned into six commonly used EEG scalp regions, namely the frontal pole, frontal, temporal, central, parietal, and occipital regions. For each predefined region $\mathcal{R}_{q}$, an attention-weighted regional feature summary and position summary are computed as
\begin{equation}
\omega_{iq}
=
\frac{\exp(\psi(\mathbf{h}_{i}^{loc}))}{\sum_{j\in\mathcal{R}_{q}}\exp(\psi(\mathbf{h}_{j}^{loc}))},
\end{equation}
\begin{equation}
\mathbf{r}_{q}=\sum_{i\in\mathcal{R}_{q}}\omega_{iq}\mathbf{h}_{i}^{loc},
\qquad
\mathbf{c}_{q}=\sum_{i\in\mathcal{R}_{q}}\omega_{iq}\mathbf{p}_{i},
\end{equation}
where $\psi(\cdot)$ is a learnable scoring function, $\mathbf{r}_{q}\in\mathbb{R}^{F_h}$ is the regional representation, and $\mathbf{c}_{q}\in\mathbb{R}^{3}$ is the corresponding regional coordinate summary.

The contribution of region $q$ to channel $i$ is then estimated by jointly considering feature relevance and spatial proximity:
\begin{equation}
\alpha_{iq}
=
\frac{\exp\!\left((\mathbf{W}_{1}\mathbf{h}_{i}^{loc})^{\top}(\mathbf{W}_{2}\mathbf{r}_{q})-\lambda\|\mathbf{p}_{i}-\mathbf{c}_{q}\|_{2}\right)}
{\sum_{q'}\exp\!\left((\mathbf{W}_{1}\mathbf{h}_{i}^{loc})^{\top}(\mathbf{W}_{2}\mathbf{r}_{q'})-\lambda\|\mathbf{p}_{i}-\mathbf{c}_{q'}\|_{2}\right)},
\end{equation}
where $\mathbf{W}_{1}$ and $\mathbf{W}_{2}$ are learnable projection matrices and $\lambda$ balances semantic similarity and spatial distance. The region-enhanced representation is obtained by
\begin{equation}
\mathbf{h}_{i}^{reg}
=
\sum_{q}\alpha_{iq}\mathbf{r}_{q}.
\end{equation}
By stacking all $\mathbf{h}_{i}^{reg}$, we obtain $\mathbf{H}^{reg}\in\mathbb{R}^{K\times F_h}$. The final spatial prior is
\begin{equation}
\mathbf{G}_{spa}
=
\mathrm{Proj}\!\left(\mathbf{H}^{loc}+\mathbf{H}^{reg}\right),
\end{equation}
where $\mathrm{Proj}(\cdot)$ maps features to $\mathbb{R}^{K\times F_{spa}}$.

\subsection{Conditional State-Space Diffusion Reconstructor}

Given the spatial prior $\mathbf{G}_{spa}$, CSDR reconstructs the target-channel EEG through conditional diffusion. Let $\mathbf{X}_{0}^{tar}\in\mathbb{R}^{K_{tar}\times L}$ denote the clean target EEG. In the forward process, Gaussian noise is gradually added as
\begin{equation}
\mathbf{X}_{t}^{tar}
=
\sqrt{\bar{\alpha}_{t}}\,\mathbf{X}_{0}^{tar}
+
\sqrt{1-\bar{\alpha}_{t}}\,\boldsymbol{\epsilon},
\qquad
\boldsymbol{\epsilon}\sim\mathcal{N}(\mathbf{0},\mathbf{I}),
\end{equation}
where $\bar{\alpha}_{t}=\prod_{i=1}^{t}(1-\beta_i)$ and $\{\beta_t\}_{t=1}^{T}$ is the variance schedule.

The reverse process learns
\begin{equation}
\begin{aligned}
p_{\theta}(\mathbf{X}_{0:T}^{tar}\mid \mathbf{X}^{obs},\mathbf{G}_{spa})
&= p(\mathbf{X}_{T}^{tar}) \\
&\quad \times \prod_{t=1}^{T}
p_{\theta}(\mathbf{X}_{t-1}^{tar}\mid \mathbf{X}_{t}^{tar},\mathbf{X}^{obs},\mathbf{G}_{spa}).
\end{aligned}
\end{equation}
This process progressively generates the missing channels under the guidance of sparse observations and topology-aware priors.

To condition the denoiser on both signal evidence and structural context, we construct a full-layout condition representation by zero-filling the missing channels in $\mathbf{X}^{obs}$ and concatenating the result with the projected spatial prior:
\begin{equation}
\mathbf{X}^{cond}
=
\mathrm{Concat}\!\left(\mathrm{ZeroFill}(\mathbf{X}^{obs}),\, \mathbf{W}_{s}\mathbf{G}_{spa}\right),
\end{equation}
where $\mathbf{W}_{s}\in\mathbb{R}^{F_{spa}\times F_c}$ is learnable. Here, $\mathrm{ZeroFill}(\mathbf{X}^{obs})$ maps the observed EEG to the full electrode layout by filling target-channel positions with zeros.

The denoiser adopts alternating temporal and channel-wise state-space modeling. Before denoising, the condition representation, noisy target state, and diffusion-step embedding are projected into a common feature space and fused to form the initial denoising state $\mathbf{z}_0\in\mathbb{R}^{K\times L\times F_d}$. Let $\mathbf{z}$ denote a generic intermediate representation in the subsequent denoising blocks. For each denoising block, temporal dynamics are first updated within each channel, after which information is propagated across channels. The temporal update is
\begin{equation}
\mathbf{z}^{tmp}
=
\mathrm{TemporalSSM}(\mathbf{z})+\mathbf{z},
\end{equation}
and the channel-wise update is
\begin{equation}
\mathbf{z}^{ch}
=
\left(
\mathrm{ChannelSSM}(\mathbf{z}^{tmp\top})+\mathbf{z}^{tmp\top}
\right)^{\top},
\end{equation}
where $(\cdot)^{\top}$ swaps the temporal and channel dimensions. The denoising network in CSDR thus serves as a conditional noise estimator parameterized by alternating temporal and channel-wise state-space blocks. Accordingly, the $n$-th denoising layer is written as
\begin{equation}
\mathbf{z}_{n}
=
\mathcal{B}_{ch}\big(\mathcal{B}_{tmp}(\mathbf{z}_{n-1})\big),
\end{equation}
where $\mathcal{B}_{tmp}$ and $\mathcal{B}_{ch}$ denote the temporal and channel-wise state-space updates, respectively. This alternating design enables the denoiser to capture long-range temporal dependencies and cross-channel interactions within a unified backbone.

During training, CSDR predicts the diffusion noise added to the target EEG. The objective is
\begin{equation}
\mathcal{L}
=
\mathbb{E}_{\mathbf{X}_{0}^{tar},\,\boldsymbol{\epsilon},\,t}
\left[
\left\|
\boldsymbol{\epsilon}
-
\boldsymbol{\epsilon}_{\theta}(\mathbf{X}_{t}^{tar},t\mid \mathbf{X}^{cond})
\right\|_{2}^{2}
\right].
\end{equation}

\subsection{Training and Inference}

TGSD is trained end-to-end under fixed observed and target channel partitions for each super-resolution factor. In each iteration, HSPE first extracts the spatial prior $\mathbf{G}_{spa}$ from $\mathbf{X}^{obs}$ and $\mathbf{P}$, after which CSDR is optimized using the noise prediction objective above.

During inference, $\mathbf{G}_{spa}$ is computed once from the sparse observations and kept fixed throughout sampling. Starting from Gaussian noise $\mathbf{X}_{T}^{tar}\sim\mathcal{N}(\mathbf{0},\mathbf{I})$, the target EEG is iteratively generated by
\begin{equation}
\begin{aligned}
\mathbf{X}_{t-1}^{tar}
&=
\frac{1}{\sqrt{\alpha_t}}
\left(
\mathbf{X}_{t}^{tar}
-
\frac{1-\alpha_t}{\sqrt{1-\bar{\alpha}_t}}
\boldsymbol{\epsilon}_{\theta}(\mathbf{X}_{t}^{tar},t\mid \mathbf{X}^{cond})
\right) \\
&\quad + \sigma_t\mathbf{z},
\end{aligned}
\end{equation}
where $\mathbf{z}\sim\mathcal{N}(\mathbf{0},\mathbf{I})$ for $t>1$ and $\mathbf{z}=\mathbf{0}$ for $t=1$. After $T$ reverse steps, the reconstructed target EEG $\hat{\mathbf{X}}_{0}^{tar}$ is obtained. The final high-resolution EEG is then formed by merging $\hat{\mathbf{X}}_{0}^{tar}$ with the observed channels over the original electrode order:
\begin{equation}
\hat{\mathbf{X}}^{HR}=\mathrm{Merge}(\mathbf{X}^{obs},\hat{\mathbf{X}}_{0}^{tar}),
\end{equation}
where $\mathrm{Merge}(\cdot)$ restores the complete channel layout.

\section{Experiments and Results}
\label{title:Experiments and results}

\begin{table*}[t]
\renewcommand{\arraystretch}{1.2}
    \begin{center}
        \small
        \caption{Reconstruction results on SEED under different spatial super-resolution factors. NMSE and PCC are reported as mean $\pm$ standard deviation. The best and second-best results in each column are highlighted in \textbf{bold} and underlined, respectively. Statistical significance: $** p < 0.01$  compared to the best baseline.}
        \label{tab:seed_recon}
        \begin{tabular*}{\linewidth}{@{\extracolsep{\fill}}ccccccc}
            \hline
            \multirow{2}{*}{Method}  & \multicolumn{2}{c}{Super-Resolution Factor 2} & \multicolumn{2}{c}{Super-Resolution Factor 4} & \multicolumn{2}{c}{Super-Resolution Factor 8} \\ 
            \cline{2-3} \cline{4-5} \cline{6-7}
            & NMSE($\downarrow$) & PCC($\uparrow$) & NMSE($\downarrow$) & PCC($\uparrow$) & NMSE($\downarrow$) & PCC($\uparrow$) \\
            \hline
            SI \cite{freeden1984spherical} & 0.684$\pm$0.000 & 0.585$\pm$0.000 & 0.703$\pm$0.000 & 0.567$\pm$0.000 & 0.755$\pm$0.000 & 0.527$\pm$0.000 \\
            EEGSR-GAN \cite{corley2018deep} & 0.514$\pm$0.002 & 0.693$\pm$0.001 & 0.597$\pm$0.002 & 0.629$\pm$0.001 & 0.652$\pm$0.001 & 0.583$\pm$0.001 \\
            Deep-CNN \cite{han2018feasibility} & 0.485$\pm$0.005 & 0.713$\pm$0.004 & 0.608$\pm$0.006 & 0.622$\pm$0.006 & 0.672$\pm$0.007 & 0.569$\pm$0.002 \\
            Deep-EEGSR \cite{tang2022deep} & 0.413$\pm$0.010 & 0.762$\pm$0.007 & 0.524$\pm$0.002 & 0.682$\pm$0.002 & 0.580$\pm$0.005 & 0.641$\pm$0.003 \\
            STAD \cite{wang2025generative} & 0.432$\pm$0.004 & 0.714$\pm$0.002 & 0.484$\pm$0.002 & 0.691$\pm$0.001 & 0.512$\pm$0.007 & 0.637$\pm$0.004 \\
            ESTformer \cite{li2025estformer} & 0.315$\pm$0.014 & 0.827$\pm$0.009 & 0.353$\pm$0.007 & 0.803$\pm$0.004 & 0.425$\pm$0.011 & 0.757$\pm$0.007 \\
            ESTformer+CNN \cite{li2025estformer} & \underline{0.277$\pm$0.014} & \underline{0.852$\pm$0.008} & \underline{0.311$\pm$0.005} & \underline{0.829$\pm$0.003} & \underline{0.377$\pm$0.009} & \underline{0.788$\pm$0.005} \\
            \hline
            TGSD & \textbf{0.237$\pm$0.008}** & \textbf{0.895$\pm$0.005}** & \textbf{0.283$\pm$0.003}** & \textbf{0.840$\pm$0.002}** & \textbf{0.337$\pm$0.009}** & \textbf{0.811$\pm$0.005}** \\
            \hline
        \end{tabular*}
    \end{center}
\end{table*}

\subsection{Experimental Setup}

\subsubsection{Datasets and Preprocessing}

We evaluate TGSD on two public EEG datasets: SEED\cite{zheng2015investigating} and PhysioNet Motor Movement/Imagery (MM/I)\cite{goldberger2000physiobank,schalk2004bci2000}, which represent emotion recognition and motor imagery scenarios, respectively. SEED contains 62-channel EEG recordings collected during affective video watching, and the preprocessed 200 Hz version is used in our experiments. PhysioNet MM/I contains 64-channel EEG recordings from 109 subjects sampled at 160 Hz. More detailed descriptions of the two datasets are provided in the supplementary material.

For preprocessing, SEED is segmented into non-overlapping 4-second windows and PhysioNet MM/I into non-overlapping 10-second windows. Each segment is normalized independently. To maintain a consistent spatial representation across datasets, all electrodes are aligned to a unified 3D coordinate system following HEAR \cite{chen2025hear}, and the resulting coordinates are used to construct the full-channel electrode layout. For both datasets, the samples are randomly split into training and test sets at a ratio of 80\%:20\%, and 10\% samples in the training set is further held out as a validation set for model selection.

\subsubsection{Spatial Super-Resolution Protocol}

Experiments are conducted under three spatial super-resolution factors, i.e., $2\times$, $4\times$, and $8\times$. For each factor, four predefined low-density electrode layouts (Case1--Case4) are adopted following \cite{tang2022deep}. The full-channel EEG serves as the high-resolution reference, while the retained electrodes form the observed channels and the remaining electrodes are treated as reconstruction targets. Training and testing are performed separately for each factor and layout, and all compared methods use the same observed/target channel partitions. The detailed electrode layouts under different super-resolution factors are provided in the supplementary material.

To simulate practical wearable EEG acquisition, we adopt channel-level missingness, where target channels remain entirely unobserved within each segment rather than being randomly missing at isolated time points. This setting matches the low-density sensing scenario considered in this work.

\subsubsection{Evaluation Metrics}

Reconstruction quality is evaluated using normalized mean squared error (NMSE) and Pearson correlation coefficient (PCC), while downstream utility is measured by classification accuracy. NMSE quantifies the normalized reconstruction error, PCC measures the correlation between reconstructed and ground-truth signals, and accuracy evaluates whether the reconstructed EEG preserves task-relevant discriminative information. For downstream evaluation, band-wise differential entropy and power spectral density features are extracted from the reconstructed EEG, and a random forest classifier with 100 trees is used to measure classification performance. Paired $t$-tests are used to assess statistical significance (* $p < 0.05$, ** $p < 0.01$).

\subsubsection{Baseline Methods}

To comprehensively validate the effectiveness of TGSD, we compare it with representative EEG spatial super-resolution methods from three categories, including traditional interpolation methods, deep learning-based reconstruction methods, and diffusion-based generative methods. Specifically, spherical spline interpolation (SI) \cite{freeden1984spherical} was adopted as a classical interpolation baseline. The deep learning-based methods include Deep-CNN \cite{han2018feasibility}, EEGSR-GAN \cite{corley2018deep}, Deep-EEGSR \cite{tang2022deep}, ESTformer \cite{li2025estformer}, and ESTformer+CNN \cite{li2025estformer}, which respectively represent CNN-based modeling, adversarial reconstruction, connectivity-aware reconstruction, transformer-based spatiotemporal modeling, and the hybrid ESTformer-CNN model proposed in the original work. In addition, STAD \cite{wang2025generative} was included as a representative diffusion-based generative baseline. These methods provide a diverse set of competitive baselines for evaluating TGSD.

\begin{table*}[t]
\renewcommand{\arraystretch}{1.2}
    \begin{center}
        \small
        \caption{Reconstruction results on PhysioNet MM/I under different spatial super-resolution factors. NMSE and PCC are reported as mean $\pm$ standard deviation. The best and second-best results in each column are highlighted in \textbf{bold} and underlined, respectively. Statistical significance: $** p < 0.01$  compared to the best baseline.}
        \label{tab:physionet_recon}
        \begin{tabular*}{\linewidth}{@{\extracolsep{\fill}}ccccccc}
            \hline
            \multirow{2}{*}{Method}  & \multicolumn{2}{c}{Super-Resolution Factor 2} & \multicolumn{2}{c}{Super-Resolution Factor 4} & \multicolumn{2}{c}{Super-Resolution Factor 8} \\ 
            \cline{2-3} \cline{4-5} \cline{6-7}
            & NMSE($\downarrow$) & PCC($\uparrow$) & NMSE($\downarrow$) & PCC($\uparrow$) & NMSE($\downarrow$) & PCC($\uparrow$) \\
            \hline
            SI \cite{freeden1984spherical} & 0.287$\pm$0.000 & 0.855$\pm$0.000 & 0.317$\pm$0.000 & 0.839$\pm$0.000 & 0.367$\pm$0.000 & 0.811$\pm$0.000 \\
            EEGSR-GAN \cite{corley2018deep} & 0.224$\pm$0.000 & 0.887$\pm$0.001 & 0.258$\pm$0.002 & 0.860$\pm$0.001 & 0.308$\pm$0.003 & 0.832$\pm$0.003 \\
            Deep-CNN \cite{han2018feasibility} & 0.206$\pm$0.019 & 0.900$\pm$0.007 & 0.231$\pm$0.021 & 0.879$\pm$0.012 & 0.276$\pm$0.044 & 0.855$\pm$0.019 \\
            Deep-EEGSR \cite{tang2022deep} & 0.165$\pm$0.001 & 0.920$\pm$0.000 & 0.177$\pm$0.001 & 0.907$\pm$0.000 & 0.214$\pm$0.002 & 0.885$\pm$0.001 \\
            STAD \cite{wang2025generative} & 0.221$\pm$0.008 & 0.893$\pm$0.004 & 0.240$\pm$0.007 & 0.884$\pm$0.003 &  0.289$\pm$0.005 &  0.824$\pm$0.002  \\
            ESTformer \cite{li2025estformer} & 0.174$\pm$0.004 & 0.908$\pm$0.002 & 0.189$\pm$0.002 & 0.900$\pm$0.001 & 0.237$\pm$0.007 & 0.872$\pm$0.004 \\
            ESTformer+CNN \cite{li2025estformer} & \underline{0.155$\pm$0.003} & \underline{0.918$\pm$0.002} & \underline{0.166$\pm$0.002} & \underline{0.912$\pm$0.001} & \underline{0.207$\pm$0.006} & \underline{0.890$\pm$0.004} \\
            \hline
            TGSD & \textbf{0.121$\pm$0.005}** & \textbf{0.942$\pm$0.005}** &  \textbf{0.143$\pm$0.002}** & \textbf{0.934$\pm$0.009}** & \textbf{0.168$\pm$0.004}** & \textbf{0.909$\pm$0.007}**\\
            \hline
        \end{tabular*}
    \end{center}
\end{table*}

\subsection{Reconstruction Performance}

Tables~\ref{tab:seed_recon} and \ref{tab:physionet_recon} report the reconstruction results on SEED and PhysioNet MM/I, respectively. TGSD achieves the best results in all reported settings, consistently obtaining the lowest NMSE and the highest PCC across both datasets and all super-resolution factors. These results indicate that TGSD improves both reconstruction accuracy and signal-level consistency under sparse-channel input.

On SEED, TGSD consistently outperforms all baselines at $2\times$, $4\times$, and $8\times$ super-resolution. Compared with the strongest baseline ESTformer+CNN, TGSD reduces NMSE from 0.277 to 0.237 at $2\times$, from 0.311 to 0.283 at $4\times$, and from 0.377 to 0.337 at $8\times$, while improving PCC from 0.852 to 0.895, from 0.829 to 0.840, and from 0.788 to 0.811, respectively. The gains over SI, EEGSR-GAN, and Deep-CNN are even larger, showing that classical interpolation and deterministic regression are insufficient for recovering complex cross-regional EEG patterns under severe channel sparsity. Moreover, the improvements over ESTformer, ESTformer+CNN, Deep-EEGSR, and STAD suggest that combining topology-aware spatial priors with conditional diffusion yields stronger structural guidance than purely discriminative reconstruction.

On PhysioNet MM/I, TGSD again ranks first across all scales. Compared with ESTformer+CNN, TGSD reduces NMSE from 0.155 to 0.121 at $2\times$, from 0.166 to 0.143 at $4\times$, and from 0.207 to 0.168 at $8\times$. The corresponding PCC is improved from 0.918 to 0.942, from 0.912 to 0.934, and from 0.890 to 0.909. Notably, ESTformer and ESTformer+CNN are already strong spatiotemporal baselines on this dataset, yet TGSD still achieves clear gains across all settings. This result indicates that explicit full-layout topology modeling and uncertainty-aware diffusion reconstruction provide complementary benefits beyond strong discriminative sequence modeling.

A common trend across both datasets is that reconstruction becomes more difficult as the super-resolution factor increases, as reflected by higher NMSE and lower PCC for all methods. Nevertheless, TGSD maintains clear advantages even under the most challenging $8\times$ setting, indicating strong robustness to severe channel missingness. This advantage can be attributed to two aspects: HSPE provides topology-aware priors over the complete electrode layout, while CSDR performs uncertainty-aware reconstruction and jointly models long-range temporal dynamics and cross-channel dependencies. Overall, these results demonstrate the effectiveness and generalizability of TGSD across different EEG paradigms and sparse-sampling conditions.

\begin{table} 
\renewcommand{\arraystretch}{1.5}
    \begin{center}
        \caption{Downstream classification accuracy (\%) on the SEED and PhysioNet MM/I datasets under different spatial super-resolution factors. Ground Truth (GT) denotes the original full-channel EEG, and LR denotes the low-density input EEG. The best and second-best reconstructed results in each column are highlighted in \textbf{bold} and underlined, respectively.}
        \label{tab:Downstream}
        \begin{tabular*}{\linewidth}{@{\extracolsep{\fill}}ccccccc}
            \hline
            Method & \multicolumn{3}{c}{SEED} & \multicolumn{3}{c}{PhysioNet MM/I} \\
            \hline
            GT & \multicolumn{3}{c}{71.5$\pm$0.6} & \multicolumn{3}{c}{88.1$\pm$0.5} \\
            \hline
            & 2x & 4x & 8x & 2x & 4x & 8x \\
            \cline{2-4} \cline{5-7}
            LR & 49.8$\pm$1.4 & 47.0$\pm$1.5 & 44.2$\pm$1.6 & 70.4$\pm$1.2 & 67.7$\pm$1.3 & 62.4$\pm$1.5\\
            STAD\cite{wang2025generative} & 54.4$\pm$1.3 & 52.5$\pm$1.4 & 46.1$\pm$1.5 & 80.7$\pm$1.1 & 77.9$\pm$1.2 & 73.1$\pm$1.3\\
            ESTformer\cite{li2025estformer} & \underline{68.9$\pm$0.9} & \underline{65.1$\pm$1.0} & \underline{60.6$\pm$1.1} & \underline{82.3$\pm$0.8} & \underline{80.2$\pm$0.9} & \underline{76.5$\pm$1.0}\\
            \hline
            TGSD & \textbf{70.5$\pm$0.7}* & \textbf{68.4$\pm$0.8}* & \textbf{62.5$\pm$0.9}* & \textbf{85.0$\pm$0.6}* & \textbf{84.7$\pm$0.7}* & \textbf{81.9$\pm$0.8}* \\
            \hline
        \end{tabular*}
    \end{center}
\end{table}

\begin{table}[t]
    \renewcommand{\arraystretch}{1.2}
    \centering
    \caption{Ablation study on SEED under the \(2\times\) super-resolution setting. The best result in each column is highlighted in \textbf{bold}.}
    \label{tab:ablation}
    \begin{tabular}{llcc}
    \hline
    \textbf{Module} & \textbf{Component Settings} & \textbf{NMSE}$\downarrow$ & \textbf{PCC}$\uparrow$ \\
    \hline
    \multirow{3}{*}{HSPE} 
    & w/o Local Topology Modeling & 0.247$\pm$0.009 & 0.883$\pm$0.006 \\
    & w/o Region-Aware Spatial Fusion & 0.261$\pm$0.011 & 0.874$\pm$0.007 \\
    & w/o HSPE & 0.273$\pm$0.012 & 0.863$\pm$0.008 \\
    \hline
    \multirow{2}{*}{CSDR} 
    & w/o TemporalSSM & 0.258$\pm$0.010 & 0.885$\pm$0.006 \\
    & w/o ChannelSSM & 0.252$\pm$0.009 & 0.886$\pm$0.005 \\
    \hline
    \multicolumn{2}{c}{TGSD} & \textbf{0.237$\pm$0.008} & \textbf{0.895$\pm$0.005} \\
    \hline
    \end{tabular}
\end{table}

\subsection{Downstream Task Performance}

Table~\ref{tab:Downstream} reports downstream classification accuracy on SEED and PhysioNet MM/I. Compared with GT, LR leads a clear performance drop on both datasets, confirming that sparse electrode sampling removes task-relevant spatial information and weakens downstream decoding.

After reconstruction, TGSD consistently achieves the best accuracy across all super-resolution factors on both datasets. On SEED, TGSD improves the accuracy from 49.8\%, 47.0\%, and 44.2\% for LR to 70.5\%, 68.4\%, and 62.5\% under $2\times$, $4\times$, and $8\times$, respectively. It also outperforms the strongest baseline ESTformer at all scales, with especially clear gains at $4\times$ and $8\times$. Notably, at $2\times$, TGSD reaches 70.5\%, which is already close to the full-channel upper bound of 71.5\%, indicating that the reconstructed EEG preserves most of the discriminative information required for emotion recognition.

A similar trend is observed on PhysioNet MM/I. TGSD improves the accuracy from 70.4\%, 67.7\%, and 62.4\% for LR to 85.0\%, 84.7\%, and 81.9\%, and consistently surpasses both STAD and ESTformer. The gains over ESTformer become more pronounced as the super-resolution factor increases, suggesting that TGSD better preserves task-relevant spatial structure under more severe channel sparsity. Combined with the reconstruction results, these findings indicate that TGSD not only improves signal-level fidelity but also recovers neural patterns that remain effective for downstream emotion recognition and motor imagery classification.

\begin{figure}
    \centering
    \includegraphics[width = 0.8\textwidth]{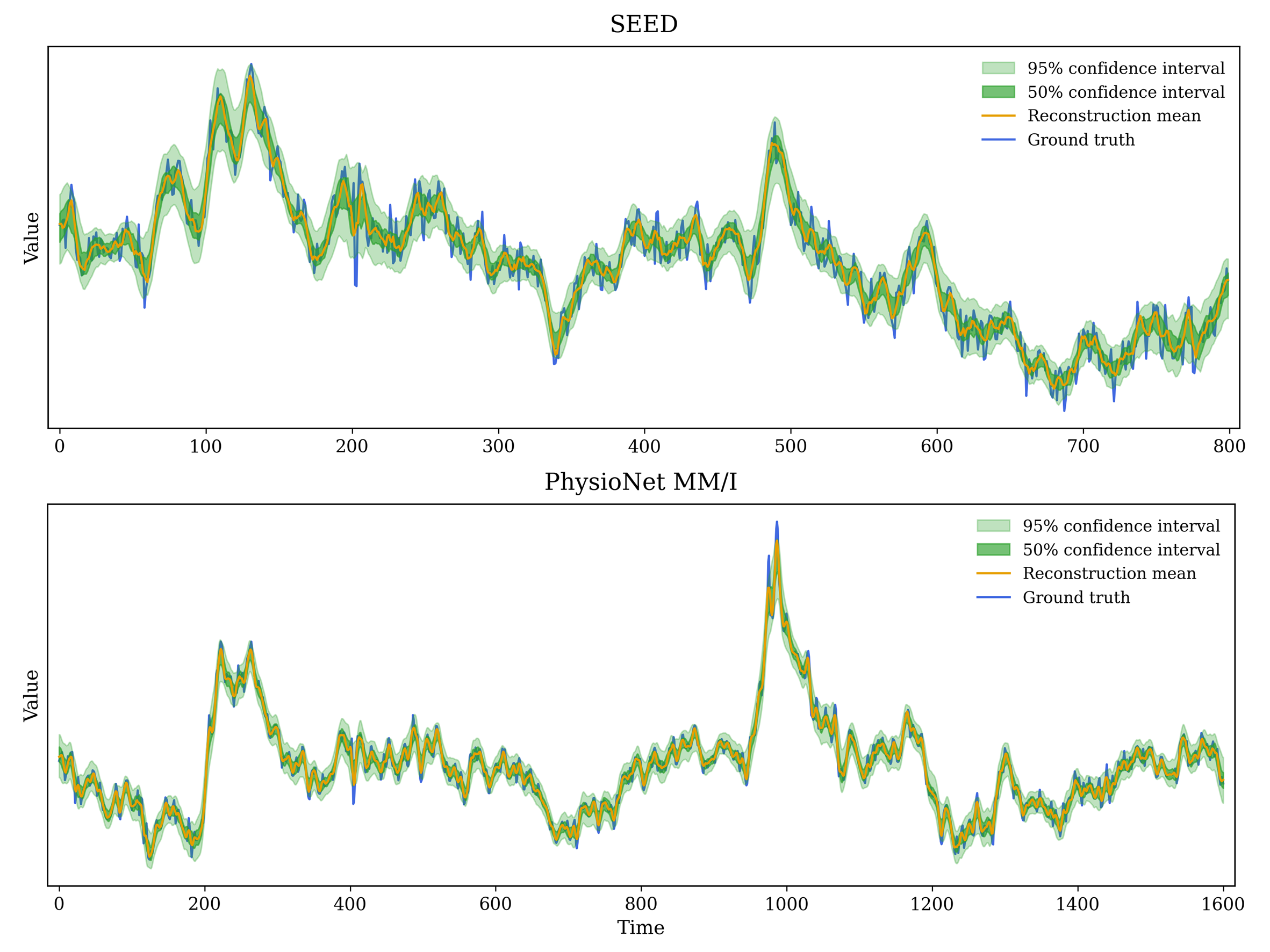}
    \caption{Time-domain visualization of probabilistic reconstruction for representative target channels under the \(2\times\) super-resolution setting on the SEED and PhysioNet MM/I datasets. The blue curve denotes the ground truth, the orange curve denotes the reconstruction mean, and the dark green and light green regions respectively indicate the 50\% and 95\% confidence intervals produced by TGSD.}
    \label{fig:time}
\end{figure}

\begin{figure}
    \centering
    \includegraphics[width = 0.8\textwidth]{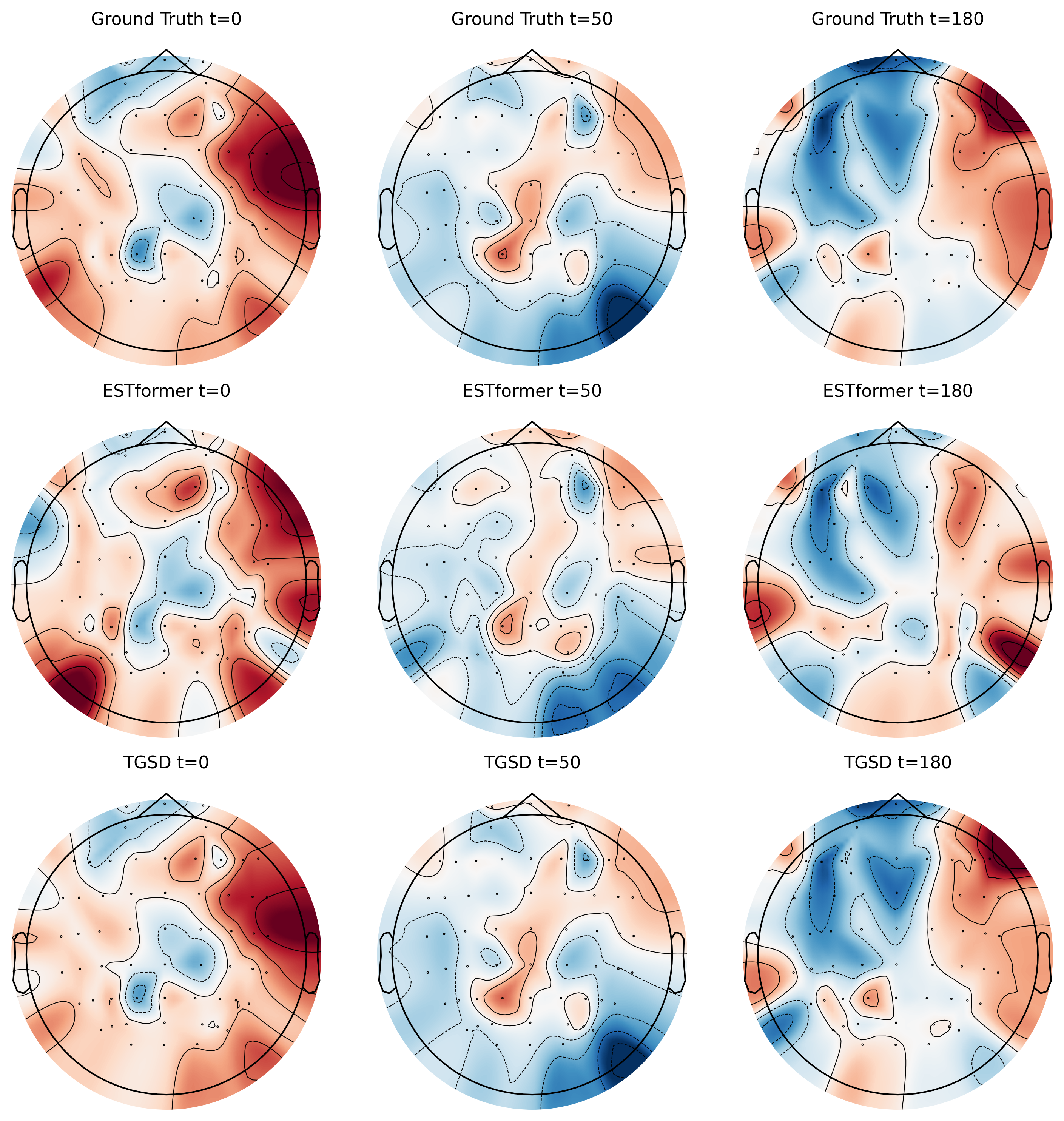}
    \caption{Topographic visualization of ground-truth EEG, ESTformer, and TGSD on the SEED dataset under the \(2\times\) super-resolution setting at three representative time steps.}
    \label{fig:topomap}
\end{figure}

\subsection{Ablation Study}

To verify the effectiveness of the key components in TGSD, we conducted an ablation study on the SEED dataset under the \( 2\times  \) spatial super-resolution setting. The results are reported in Table~\ref{tab:ablation}. We mainly examine the roles of local topology modeling and region-aware spatial fusion in HSPE, as well as the contributions of temporal state-space modeling and channel state-space modeling in CSDR. 

From the results related to HSPE, removing local topology modeling increases the NMSE from 0.237 to 0.247 and reduces the PCC from 0.895 to 0.883, indicating that local spatial propagation based on electrode adjacency is important for target-channel reconstruction. When region-aware spatial fusion is further removed, the performance degrades more noticeably, with the NMSE and PCC degrading to 0.261 and 0.874, respectively. This suggests that local geometric relations alone are insufficient to fully characterize cross-regional spatial dependencies in EEG, while region-level contextual information provides effective complementary structural priors for reconstruction. When the entire HSPE is removed, the performance further drops to an NMSE of 0.273 and a PCC of 0.863, demonstrating that hierarchical spatial prior encoding is critical for improving reconstruction quality.

For the ablation results of CSDR, removing either TemporalSSM or ChannelSSM leads to degraded performance. Specifically, without TemporalSSM, the NMSE and PCC deteriorate to 0.258 and 0.885, respectively; without ChannelSSM, the corresponding results are 0.252 and 0.886. These results indicate that EEG spatial super-resolution depends not only on modeling long-range temporal dynamics within individual channels, but also on information propagation and interaction across channels. In comparison, the complete alternating temporal--channel state-space modeling strategy can jointly capture temporal and spatial dependencies more effectively, thereby yielding the best reconstruction performance.

Overall, the full TGSD achieves the lowest NMSE and the highest PCC among all settings, which confirms the effectiveness of the key designs in HSPE and CSDR for EEG spatial super-resolution.

\subsection{Visualization and Qualitative Analysis}

To illustrate the probabilistic reconstruction capability of TGSD, we visualize representative target-channel results under the \(2\times\) setting in Figs.~\ref{fig:time} and \ref{fig:topomap}. The time-domain plot highlights both the reconstruction mean and the associated uncertainty, which cannot be provided by deterministic regression models.

As shown in Fig.~\ref{fig:time}, the reconstruction mean closely follows the ground truth on both SEED and PhysioNet MM/I, while the 50\% and 95\% confidence intervals cover most waveform variations. The uncertainty bands remain narrower in stable regions and widen around sharp peaks and transitions, reflecting higher reconstruction ambiguity in these segments.

Fig.~\ref{fig:topomap} further shows that TGSD produces scalp topographies more consistent with the ground truth than ESTformer on SEED under the \(2\times\) setting, with more accurate activation regions and spatial transitions. Additional topographic visualizations on the PhysioNet MM/I dataset are provided in the supplementary material (Fig.S3). These qualitative results are consistent with the quantitative improvements of TGSD.

\section{Conclusion}
\label{title:Conclusion}

In this paper, we proposed TGSD, a topology-guided state-space diffusion framework for EEG spatial super-resolution. By combining a Hierarchical Spatial Prior Encoder with a Conditional State-Space Diffusion Reconstructor, TGSD leverages full-layout topology-aware spatial priors together with probabilistic generation and temporal-channel dependency modeling to recover missing-channel EEG from sparse observations. Experimental results on the SEED and PhysioNet MM/I datasets demonstrated that TGSD consistently outperformed representative baselines under multiple spatial upsampling settings in both reconstruction fidelity and downstream classification performance. These findings indicate that topology-guided conditional diffusion provides an effective solution for enhancing low-density EEG sensing and holds promise for wearable and IoT-based neural monitoring applications. While the current study focuses on offline reconstruction under predefined electrode layouts, future work will investigate more diverse recording conditions, cross-layout generalization, and lightweight model design for more practical deployment in resource-constrained scenarios.

\section*{Acknowledegment}

This work was supported by The Hong Kong Polytechnic University Start-up Fund (Project ID: P0053210), The Hong Kong Polytechnic University Faculty Reserve Fund (Project ID: P0053738), an internal grant from The Hong Kong Polytechnic University (Project ID: P0048377), The Hong Kong Polytechnic University Departmental Collaborative Research Fund (Project ID: P0056428), The Hong Kong Polytechnic University Collaborative Research with World-leading Research Groups Fund (Project ID: P0058097) and Research Grants Council Collaborative Research Fund (Project ID: C5033-24G).

\bibliographystyle{unsrt}  
\bibliography{references}  

\section{Supplementary Materials}

\setcounter{figure}{0}
\setcounter{table}{0}
\renewcommand{\thefigure}{S\arabic{figure}}
\renewcommand{\thetable}{S\arabic{table}}

\subsection{Detailed Dataset Description}

\subsubsection{SEED Dataset}

The SEED dataset is a public benchmark for EEG-based emotion analysis \cite{zheng2015investigating}. It contains EEG recordings collected from 15 subjects while they watched film clips designed to elicit three affective states, namely positive, neutral, and negative emotions. The signals were acquired using a 62-channel EEG system arranged according to the international 10--20 electrode placement standard. In this work, we used the officially released preprocessed version of SEED, in which the recordings had been downsampled to 200 Hz and standard artifact handling procedures had already been performed.

Following a commonly adopted setting for EEG reconstruction, the continuous recordings were segmented into non-overlapping 4-second windows. Accordingly, each EEG segment contains 62 channels and 800 temporal samples. These segmented samples were used to construct sparse-to-dense EEG reconstruction pairs under different spatial super-resolution settings.

\subsubsection{PhysioNet MM/I Dataset}

The PhysioNet Motor Movement/Imagery (MM/I) dataset is a widely used public EEG dataset for motor execution and motor imagery analysis \cite{goldberger2000physiobank,schalk2004bci2000}. It includes EEG recordings from 109 subjects acquired with 64 electrodes following the international 10--10 montage. The signals were recorded at 160 Hz while the subjects performed or imagined several motor tasks, including left-fist movement, right-fist movement, both-fists movement, and both-feet movement.

In our experiments, the continuous EEG recordings were divided into non-overlapping 10-second windows. Therefore, each sample contains 64 channels and 1600 temporal points. Compared with SEED, this dataset represents a substantially different EEG application scenario, which enables us to evaluate the robustness and generalization ability of TGSD across distinct neural paradigms.

\subsubsection{Preprocessing}

For both datasets, each segmented EEG sample was normalized independently before being fed into the model. This sample-wise normalization reduces amplitude variation across different segments and subjects, and improves the stability of model training. Since the goal of this work is EEG spatial super-resolution under sparse-channel acquisition, no additional point-wise masking was introduced during preprocessing. Instead, missingness was imposed at the channel level, where an entire target channel remained unobserved throughout the whole segment.

To maintain a consistent spatial representation across datasets with different channel configurations, all electrodes were aligned to a unified 3D coordinate system following HEAR \cite{chen2025hear}. Specifically, each channel was assigned a fixed 3D coordinate according to its standard scalp position, and these coordinates were used to define the full-layout electrode topology for spatial prior learning. Under each spatial super-resolution factor, the complete channel set was partitioned into observed and target subsets according to the predefined sparse layouts described in the main paper.

After segmentation and normalization, the samples in each dataset were randomly divided into training and test subsets with a ratio of 80\%:20\%. In addition, 10\% of the training data were further held out as a validation set for model selection and hyperparameter tuning. The same data split protocol was applied to all compared methods to ensure a fair evaluation.

\subsection{Electrode Layouts for Different Super-Resolution Factors}

To evaluate EEG spatial super-resolution under different sparse acquisition conditions, we considered three super-resolution factors, namely \(2\times\), \(4\times\), and \(8\times\), on both the SEED and PhysioNet MM/I datasets. For each factor, four predefined sparse electrode layouts (Case1--Case4) were constructed, following the evaluation protocol adopted in the main paper. Under each layout, a subset of electrodes was retained as the observed channels, while the remaining electrodes were treated as target channels to be reconstructed.

\begin{figure}
    \centering
    \includegraphics[width=0.9\textwidth]{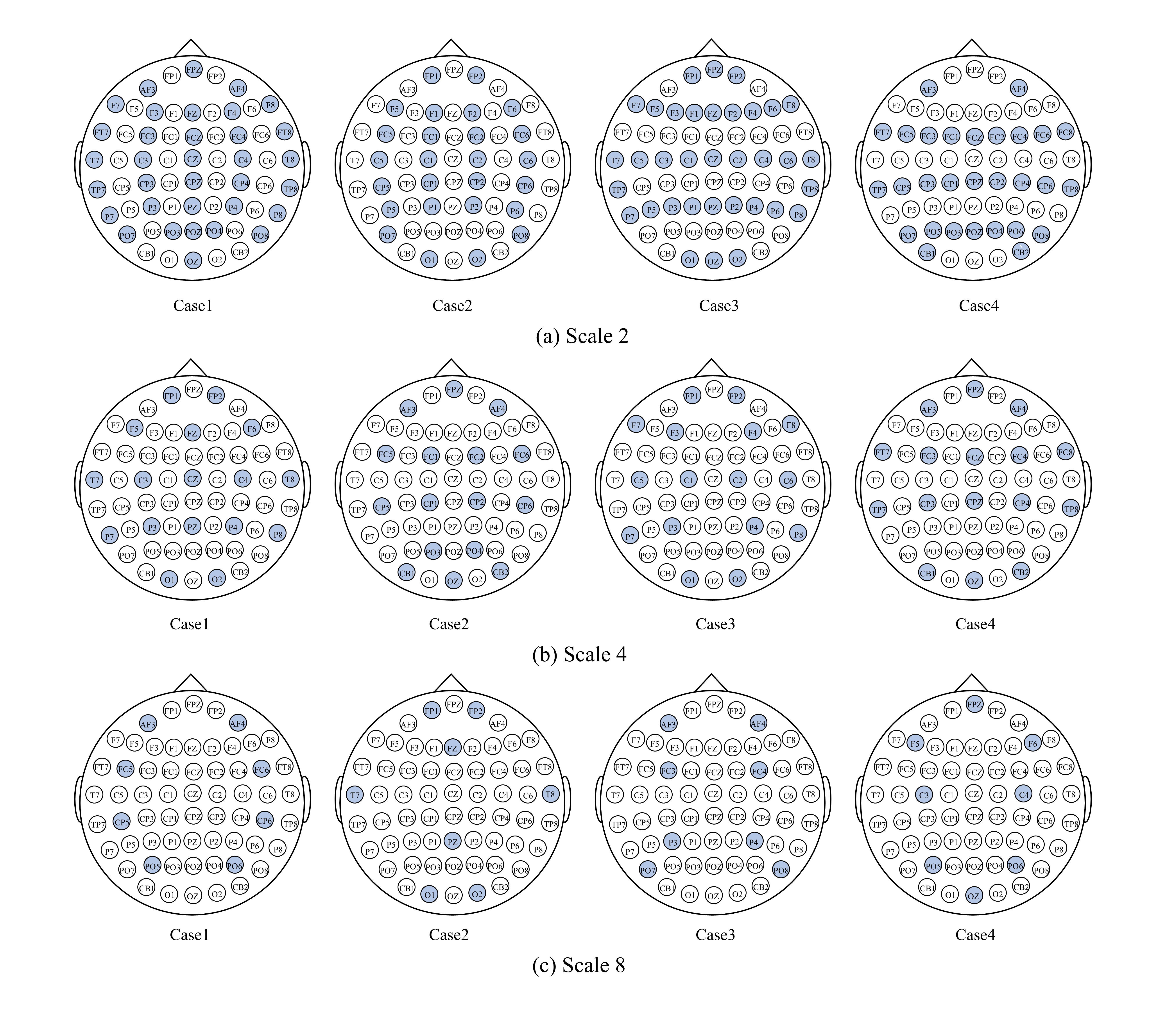}
    \caption{Predefined sparse electrode layouts for the SEED dataset under three spatial super-resolution factors: (a) \(2\times\), (b) \(4\times\), and (c) \(8\times\). In each case, blue electrodes denote observed channels and white electrodes denote target channels to be reconstructed.}
    \label{fig:seed}
\end{figure}

\begin{figure}
    \centering
    \includegraphics[width=0.9\textwidth]{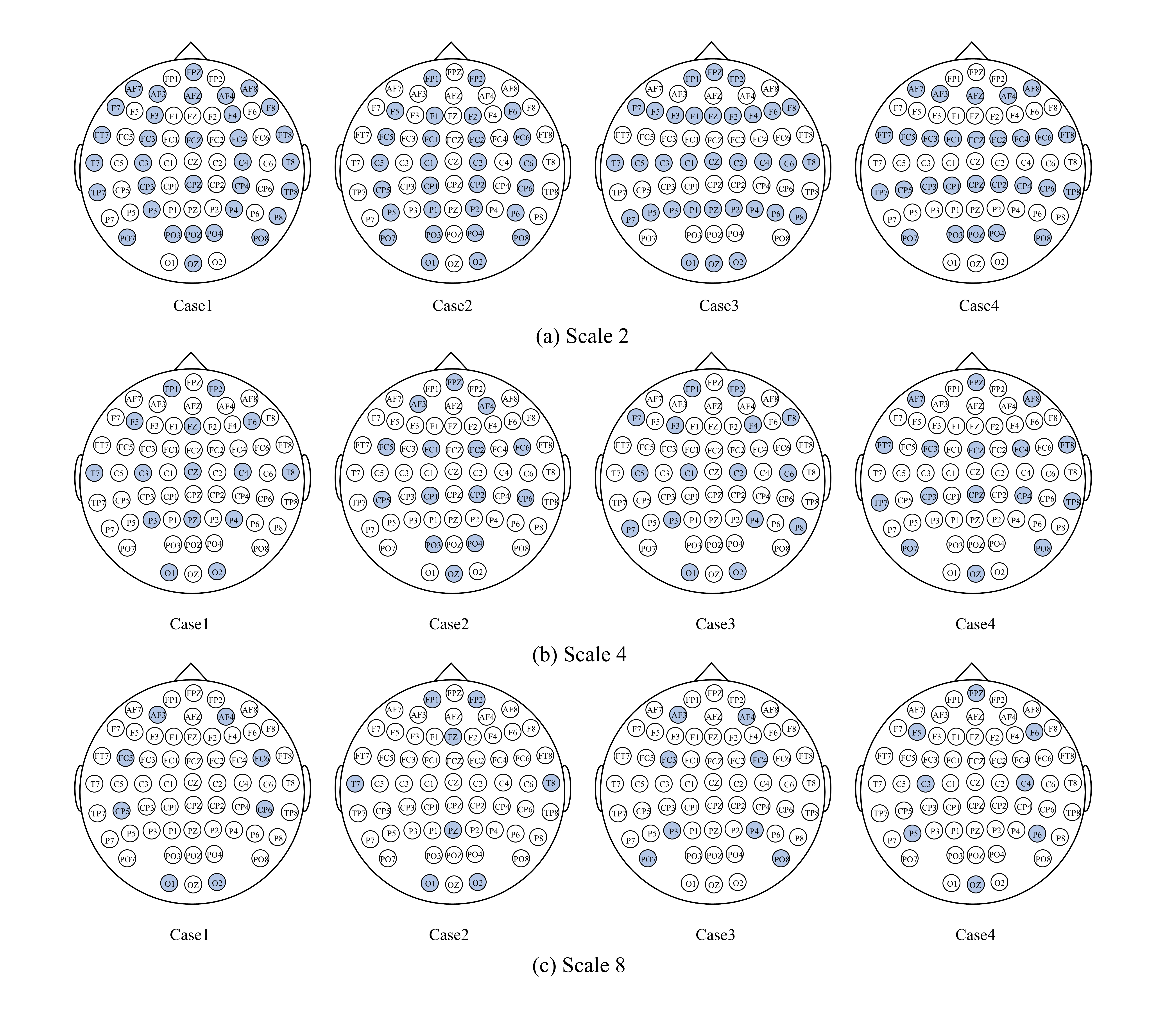}
    \caption{Predefined sparse electrode layouts for the PhysioNet MM/I dataset under three spatial super-resolution factors: (a) \(2\times\), (b) \(4\times\), and (c) \(8\times\). In each case, blue electrodes denote observed channels and white electrodes denote target channels to be reconstructed.}
    \label{fig:mmi}
\end{figure}

Figs.~\ref{fig:seed} and \ref{fig:mmi} present the electrode layouts used for the SEED and PhysioNet MM/I datasets, respectively. In each subfigure, the blue electrodes denote the observed channels available to the model as low-density input, whereas the white electrodes indicate the target channels to be reconstructed. For a given dataset, the layouts are organized by scale factor, with four different channel-retention cases shown for each factor.

The use of multiple sparse layouts under each super-resolution factor serves two important purposes. First, it reduces the possibility that the reported performance is biased toward a particular favorable channel subset. Second, it better reflects practical wearable and low-density EEG scenarios, in which different devices may retain different electrode configurations. By evaluating all compared methods under the same predefined layouts, the experimental setting provides a fair and robust assessment of reconstruction performance across varying degrees of channel sparsity.

As the super-resolution factor increases from \(2\times\) to \(8\times\), the number of observed channels decreases and the reconstruction task becomes more challenging. In particular, the \(2\times\) setting corresponds to relatively moderate channel sparsity, whereas the \(8\times\) setting represents a substantially more difficult low-density acquisition scenario. The four cases under each factor further introduce layout diversity, allowing us to examine whether a reconstruction method can generalize across different spatial sampling patterns rather than overfitting to a single electrode arrangement.

For all experiments reported in the main paper, the same observed/target channel partitions were used for training and testing within each dataset, scale factor, and case. The quantitative results shown in the main paper were obtained by averaging across the four predefined cases for each super-resolution factor.

\subsection{Implementation Details}

\subsubsection{Training Configuration}

All experiments were implemented in PyTorch and trained using the Adam optimizer. The initial learning rate was set to \(2\times10^{-4}\), and the model was trained for 300{,}000 iterations. Model selection was performed on the validation set, and the checkpoint with the best validation performance was used for final testing.

For diffusion training, the number of diffusion steps was set to \(T=200\). We adopted a linear variance schedule with \(\beta_1=10^{-4}\) and \(\beta_T=0.02\). During training, the diffusion step was uniformly sampled from \(\{1,\dots,T\}\), and the model was optimized using the standard noise-prediction objective described in the main paper.

To ensure a fair comparison across different datasets and super-resolution settings, the same training protocol and optimization strategy were used for all experiments unless otherwise specified.

\subsubsection{Model Hyperparameters}

The denoising network in CSDR was composed of 8 alternating temporal and channel-wise state-space blocks. The hidden feature dimension was set to 64, and the diffusion-step embedding dimension was set to 128. Unless otherwise noted, all intermediate feature projections in HSPE and CSDR were configured using the same hidden dimensionality for stable optimization and implementation simplicity.

For HSPE, the local topology graph was constructed based on Euclidean distances between electrode coordinates using a \(k\)-nearest-neighbor strategy. The number of neighbors and the associated spatial parameters were selected on the validation set and then kept fixed across all experiments. In the region-aware spatial fusion module, the six predefined scalp regions described in the main paper were used consistently for both datasets.

For CSDR, the target-channel reconstruction process was conditioned on the sparse observations and the learned topology-aware spatial prior. The same denoising depth and feature dimensions were used for all super-resolution factors, which allows the comparison across settings to focus on the effect of channel sparsity rather than architecture changes.

\subsubsection{Inference Details}

During inference, the spatial prior produced by HSPE was computed once from the observed EEG channels and the corresponding electrode coordinates, and then reused throughout the reverse diffusion process. Starting from Gaussian noise, the target-channel EEG was progressively reconstructed over \(T=200\) reverse steps. The final high-resolution EEG was obtained by merging the reconstructed target channels with the observed channels according to the original electrode ordering. Table~\ref{tab:impl_hyper} summarizes the main implementation hyperparameters used in our experiments. All experiments were conducted on a workstation equipped with an NVIDIA RTX-4090 GPU.

\begin{table}[t]
\renewcommand{\arraystretch}{1.2}
\centering
\caption{Main implementation hyperparameters of TGSD.}
\label{tab:impl_hyper}
\begin{tabular}{lc}
\hline
\textbf{Hyperparameter} & \textbf{Value} \\
\hline
Optimizer & Adam \\
Initial learning rate & \(2\times10^{-4}\) \\
Training iterations & 200,000 \\
Diffusion steps \(T\) & 200 \\
Noise schedule & Linear \\
\(\beta_1\) & \(10^{-4}\) \\
\(\beta_T\) & \(0.02\) \\
Number of denoising blocks & 8 \\
Hidden feature dimension & 64 \\
Diffusion embedding dimension & 128 \\
Number of scalp regions & 6 \\
\hline
\end{tabular}
\end{table}

\subsection{Topographic visualization}

\begin{figure}
    \centering
    \includegraphics[width=0.8\textwidth]{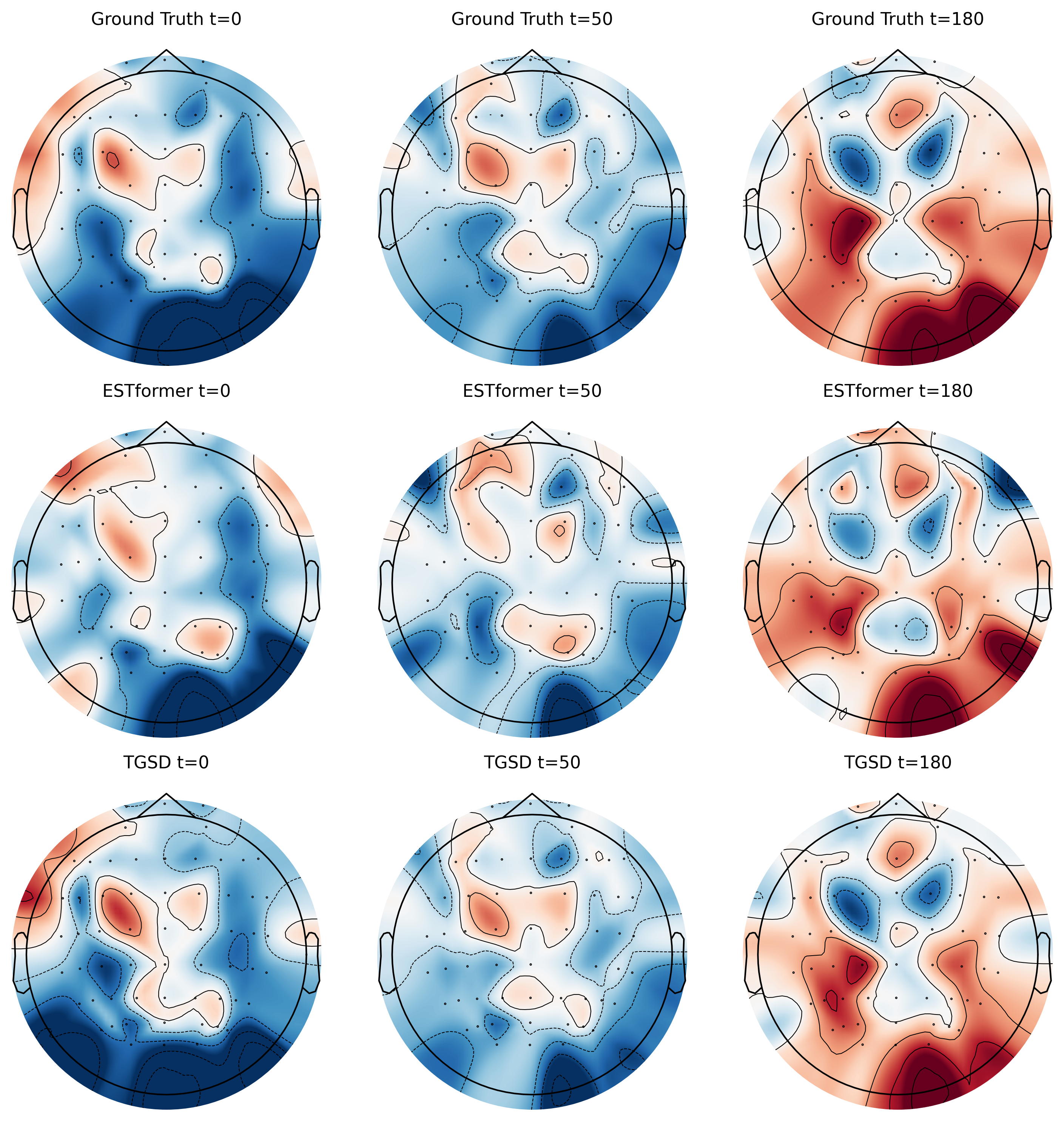}
    \caption{Topographic visualization of ground-truth EEG, ESTformer, and TGSD on the PhysioNet MM/I dataset under the \(2\times\) super-resolution setting at three representative time steps.}
    \label{fig:topomapmmi}
\end{figure}

\end{document}